*Agentic AI for Scaling Diagnosis and Care in Neurodegenerative Disease*

*Andrew G. Breithaupt[1], Michael Weiner[2], Alice Tang[3,4], Katherine L. Possin[5,6], Marina Sirota[4], James Lah[1], Allan I. Levey[1], Pascal Van Hentenryck[7], Reza Zandehshahvar[7], Marilu Luisa Gorno-Tempini[5,6], Joseph Giorgio[8], Jingshen Wang[9], Andreas M. Rauschecker[2,10], Howard J. Rosen[5,6], Rachel L. Nosheny[2,11], Bruce L. Miller[5,6], Pedro Pinheiro-Chagas[5]*

[1]Goizueta Brain Health Institute, Emory University, Atlanta, GA

[2]Department of Radiology and Biomedical Imaging, University of California, San Francisco, San Francisco, CA

[3]School of Medicine, University of California, San Francisco, San Francisco, CA

[4]Bakar Computational Health Sciences Institute, University of California, San Francisco, San Francisco, CA

[5]Memory and Aging Center, Department of Neurology, University of California, San Francisco, CA

[6]Department of Neurology, Weill Institute for Neuroscience, University of California, San Francisco, San Francisco, CA

[7]NSF AI Institute for Advances in Optimization (AI4OPT), Georgia Institute of Technology, Atlanta, GA

[8]Department of Neuroscience, University of California, Berkeley, Berkeley, CA

[9]Division of Biostatistics, University of California, Berkeley, Berkeley, CA

[10]Center for Intelligent Imaging (ci2), Department of Radiology and Biomedical Imaging, University of California, San Francisco, San Francisco, CA

[11]Department of Psychiatry and Behavioral Sciences, University of California, San Francisco, San Francisco, CA

## Abstract

United States healthcare systems are struggling to meet the growing demand for neurological care, particularly in Alzheimer's disease and related dementias (ADRD). Generative AI built on language models (LLMs) now enables agentic AI systems that can enhance clinician capabilities to approach specialist-level assessment and decision-making in ADRD care at scale. This article presents a comprehensive six-phase roadmap for responsible design and integration of such systems into ADRD care: (1) high-quality standardized data collection across modalities; (2) decision support; (3) clinical integration enhancing workflows; (4) rigorous validation and monitoring protocols; (5) continuous learning through clinical feedback; and (6) robust ethics and risk management frameworks. This human centered approach optimizes clinicians' capabilities in comprehensive data collection, interpretation of complex clinical information, and timely application of relevant medical knowledge while prioritizing patient safety, healthcare equity, and transparency. Though focused on ADRD, these principles offer broad applicability across medical specialties facing similar systemic challenges.

## Introduction

The gap between the demand for neurological care and the number of neurology providers continues to grow,[1] leading to long wait times, delayed diagnoses, and increased emergency department visits.[2,3] Neurological diagnoses are difficult to make for a wide variety of reasons ranging from the great variability and complexity in patient presentations to the extensive time and knowledge required to effectively collect and interpret necessary data. These challenges are amplified by limited office visit time, exponentially growing data complexity, and an overwhelming volume of medical literature and research opportunities that exceed any individual clinician's capacity to master.

Simultaneously, artificial intelligence (AI) is rapidly advancing and has entered a new era through generative AI built on large language models (LLMs). A growing number of studies have demonstrated that LLM-based systems encode clinical knowledge,[4] can achieve expert-level medical question answering,[5] integrate multimodal patient data,[6] are capable of empathetic history taking through conversational agents,[7] and enhance clinical decision-making.[8] LLMs have enabled agentic AI systems, architectures in which one or more LLMs coordinate interacting components such as traditional discriminative AI models and tools to perform complex, multistep tasks (See Box 1).[9,10] These systems can search medical databases, analyze structured and unstructured patient data, and connect to electronic health records. While much research has focused on AI's potential to identify patterns beyond human perception (predictive AI), clinical implementation remains challenging because clinicians cannot readily verify insights they cannot detect, nor can these systems naturally complement clinical workflows. However, agentic AI systems offer more immediately practical applications by enhancing clinicians' capabilities in three critical areas where healthcare providers increasingly struggle: comprehensive data collection, interpretation of increasingly complex multimodal information, and timely, context appropriate application of relevant medical knowledge.

Currently, we can deploy domain-specific AI assistants for individual functions, each optimized for its particular role in data collection, interpretation, or knowledge retrieval; however, the future vision involves an integrated agentic AI system capable of seamlessly coordinating these functions within a continuously learning healthcare framework. Such systems would leverage every patient encounter while incorporating the latest clinical evidence.

Still, such a learning system must be intentionally molded to benefit diverse patient populations, support both patients and providers without imposing new burdens, and be safely implemented in any clinic setting. Recent concerns around generative and agentic AI have emphasized the need for strict boundaries around autonomy and for maintaining robust human oversight in high-stakes medical contexts.[11] This is particularly important in ADRD, where vulnerable populations, shifting decision-making capacity, and high-stakes diagnostic uncertainty could magnify risk. Therefore, we advocate for constrained agentic AI systems in which clinicians remain in the loop, supported by rigorous evaluation and continuous monitoring using clinically meaningful outcomes.

In this perspective article, we provide a roadmap for how responsible agentic AI integration can be accomplished in the field of ADRD in the United States and scale specialist-level clinical care. Our roadmap consists of six interconnected phases: (1) high-quality standardized data collection, (2) AI decision support development, (3) clinical integration and workflow, (4) validation and monitoring, (5) continuous learning, and (6) ethics and risk management. While our roadmap presents clinical integration as a distinct phase, implementation success requires designing all components with clinical workflows in mind from the outset. Therefore, integration considerations are intentionally woven throughout this paper. We hope that this framework can be useful for anyone in the field of neurology or other specialties, encouraging exploration into how agentic AI may aid their patients and fellow providers in ways previously not possible.

| Box 1: Moving from Single LLMs to Agentic AI |
| --- |
| Agentic AI refers to compound systems in which large language models (LLMs) and other ML models are orchestrated through scaffolding architectures that enable code execution, search, memory, and iterative refinement, fundamentally differing from single-model approaches. While single LLMs have demonstrated impressive performance on medical licensing examinations and diagnostic benchmarks,[4,5] they are not able to execute diverse tasks in the real world. In addition, several critical limitations constrain their clinical utility, each of which agentic approaches can potentially address. First, LLMs exhibit "jagged intelligence" with gaps between strengths and weaknesses;[12] agentic scaffolding can close these gaps through systematic planning, context engineering, and verification.[9,10] Second, hallucinations persist despite improvements, particularly when models lack empirical grounding; agentic systems can enforce scientific grounding by retrieving and cross-checking outputs against trusted medical sources.[13] Single LLMs lack mechanisms for continual learning, causing their knowledge to lag behind evolving medical evidence;[14] agents can access up-to-date external sources such as FDA approvals, clinical trials, and treatment guidelines at inference time.[13] Fourth, limited context windows and sample inefficiency restrict how much clinical information models can use;[15]agentic architectures can address these constraints by integrating massive, compressed external knowledge stored in vector databases and knowledge graphs through retrieval-augmented generation (RAG).[16] Fifth, LLMs lack calibrated statistical metrics to quantify uncertainty during prediction, limiting trust; compound systems can integrate traditional machine learning models that provide quantitative, interpretable outputs.[17] |

**Table 1: Agentic AI Integration Roadmap for ADRD Care**

| Phase | Rationale & Objectives | Implementation Strategies | Potential Key Stakeholders | Expected Impact |
|---|---|---|---|---|
| **High-Quality, Standardized Data Collection** | • Build infrastructure to collect standardized, high quality, multi-modal clinical data<br>• Proper curation according to FAIR principles (Findable, Accessible, Interoperable, and Reusable)<br>• Establish structured formats for data currently lacking standardization<br>• Ensure data collection is accessible and easy for diverse patient populations and clinic settings | **Near-term:**<br>• Encourage and leverage ambient scribes for high quality documentation<br>• Validate implementation and effectiveness of AI enabled data collection in diverse settings (digital cognitive assessments, voice-based conversational agents for patient history over the phone, etc.)<br>• Establish multi-institutional data standards across all modalities<br><br>**Long-term:**<br>• Create repository of standardized, multimodal ADRD datasets accessible in the clinic setting<br>• Integrate passive monitoring from digital biomarkers in clinical care | • National organizations (NIH, CMS)<br>• Professional societies<br>• Research institutions<br>• Clinicians (dementia specialists, primary care providers (PCPs), neuropsychologists)<br>• Patient advocacy groups<br>• Domain experts for each data type (neuroimaging, biomarkers, etc.)<br>• Health informatics specialists | • Comprehensive data collection that reduces patient and clinician burden<br>• Standardized assessment across care settings<br>• Data prioritization across institutions evolves with the latest evidence (e.g., implementation of more accurate, evidence-based biomarkers) |
| **AI Decision Support Development** | • Augment clinician capabilities through clear presentation and interpretation of complex multimodal data using discriminative and generative models<br>• Enable explainable diagnostic reasoning that clinicians can verify and | **Near-term:**<br>• Train models to interpret ADRD specific multimodal data<br>• Develop LLM-based systems capable of explaining decision support recommendations mimicking clinician decision making<br>• Use methods such as RAG to ground decision support in latest evidence base<br><br>**Long-term** | • AI researchers<br>• Clinicians (dementia specialists, PCPs, neuropsychologists)<br>• Health equity researchers<br>• Implementation scientists<br>• Payers/insurers<br>• Healthcare economists | • Increase early and accurate diagnosis in primary care and specialty settings, decreasing diagnostic disparities<br>• Interpretable decision support<br>• Personalized care recommendations |
| | trust, grounded in the latest evidence base<br>• Support personalized management decisions including complex therapy eligibility criteria | • Validate implementation and effectiveness of AI decision support across diverse clinical settings and patient populations served, from academic specialty centers to resource-limited primary care practices | • Policy Makers | • Increased identification of clinical trial candidates |
| **Clinical Integration & Workflow** | • Address barriers to health technology adoption in clinical settings<br>• Alleviate clinician burnout, designing approaches that enhance rather than disrupt existing workflows<br>• Avoid patient/caregiver burden<br>• Maintain meaningful clinician oversight while maximizing automation benefits | **Near-term:**<br>• Study clinician-AI interaction to optimize clinical utility/safety and implementation in both specialty and primary care settings (mitigating underutilization and overreliance)<br>• Build Electronic Health Record (EHR) compatible tools reducing documentation burden and clicks<br>• Create patient-AI interfaces that enhance accessibility and reduce burden<br>• Develop implementation metrics that assess clinical value and workflow efficiency<br><br>**Long-term:**<br>• Develop AI assistants that can adapt interaction/explanations to clinician level of expertise<br>• Build systems for automated follow-up that maintain human oversight<br>• Implement training programs to optimize clinician-AI collaboration | • Health system leaders<br>• Clinical operations staff<br>• Frontline clinicians<br>• EHR vendors<br>• Implementation scientists<br>• Payers/insurers<br>• Healthcare economists<br>• Policy Makers | • Time savings allowing more meaningful patient-provider interactions<br>• Enhanced clinical decision making with evidence-based care becoming the path of least resistance<br>• Tools easily implemented in any care setting regardless of available resources<br>• Improved clinician satisfaction and retention<br>• AI systems with practical clinician oversight |
| **Validation, Monitoring & Governance** | • Ensure patient safety<br>• Prevent model drift & degradation<br>• Identify & mitigate biases<br>• Inform regulatory framework | **Near-term:**<br>• Establish validation protocols using FAVES framework (Fair, Appropriate, Valid, Effective, Safe)<br>• Create monitoring dashboards for model performance using benchmarks and specialist assessments | • National organizations<br>• Professional societies<br>• Dementia specialists and neuropsychologists | • Consistent AI performance across diverse clinical and geographic settings<br>• Regulatory approval pathways for AI tools |
| | • Ensure cost effectiveness and high value care | **Long-term:**<br>• Develop neuropathology and biomarker correlation where feasible<br>• Establish independent oversight committees | • Regulatory bodies<br>• Patient advocacy groups<br>• Model governance bodies<br>• Policy Makers | • Early detection of model performance degradation<br>• Technical quality assurance standards for clinical AI |
| **Continuous Learning & Improvement** | • Systematically integrate clinician feedback, patient outcomes, and emerging research to refine models while preventing quality degradation<br>• Ensure models continue to address evolving real-world needs<br>• Ensure only high-quality data is used for model updates<br>• Ensure community engagement | **Near-term:**<br>• Establish passive and voluntarily active feedback mechanisms for clinicians, and develop protocols for curating feedback<br>• Create patient outcome tracking systems<br>• Build real-time literature monitoring systems to update RAG database with human oversight<br>• Establish Community Scientific Partnership Boards<br><br>**Long-term:**<br>• Implement semi-open models with robust governance, including independent oversight and model update protocols<br>• Create clinician communities for model improvement | • Clinical research community<br>• Individuals living with dementia and their families and caregivers<br>• Model governance bodies<br>• Health informatics specialists | • Models that improve with use<br>• Rapid incorporation of new evidence<br>• Knowledge sharing across institutions<br>• Adaptation to evolving clinical practices |
| **Ethics & Risk Management** | • Adopt and advance ethical guidelines to identify and mitigate patient harm as the greatest risk<br>• Ensure fairness, privacy, explainability, and clear accountability<br>• Protect healthcare workers' rights as well | **Near-term:**<br>• Create efficient adverse event reporting systems<br>• Establish robust privacy safeguards<br>• Ensure transparency in model limitations and potential biases<br>• Design redundant safety systems<br><br>**Long-term:**<br>• Develop ADRD specific Health AI Ethics committee | • ADRD specific ethics committee<br>• Legal advisors<br>• Professional societies<br>• Patient advocacy groups<br>• Healthcare ethicists<br>• Policy Makers | • Equitable access to advanced care<br>• Accountability mechanisms that assign appropriate responsibility among providers, developers, and institutions<br>• Protected patient autonomy<br>• Transparent AI deployment |
| | | • Create AI ethics certification standards for ADRD clinical applications<br>• Establish liability and indemnification frameworks with legal and regulatory input | | • Established remediation pathways when harm occurs |

*This roadmap conceptualizes a continuous learning healthcare system where insights from monitoring and implementation (Components 3-6) feed back into improving data collection standards and model development (Components 1-2), creating an ongoing cycle of improvement.*

## A crisis in the field of ADRD

As the U.S. population ages, the number of people with Alzheimer's disease and related dementias (ADRD) is rapidly increasing,[18,19] with US cases projected to double by 2060 to 1 million annually.[20] Early detection and precise diagnosis are crucial for proper access to quality care, eligibility for new and emerging disease modifying therapies, and enabling research to advance the field. Unfortunately, more than 50% of dementia diagnoses are delayed until moderate or advanced stages in primary care[21,22] with greater delays among racial and ethnic minorities.[23]

There are many challenges with diagnosing neurodegenerative diseases. Patients usually go to their primary care providers (PCPs) first, but most PCPs cite lack of time and confidence with ADRD diagnosis.[24,25] Referrals to dementia specialists face significant barriers as the demand greatly exceeds the available specialist supply with the gap continuing to widen,[1,26,27] and the average wait times projected to exceed 40 months by 2027 with rural areas facing three times longer delays.[28] Beyond access issues, diagnosis is challenging, requiring time-consuming data collection and nuanced interpretation.

After diagnosis, challenges with management are growing as new therapeutics with intensive requirements for monitoring are emerging.[29,30] Many PCPs and general neurologists struggle to keep pace with the rapid advances in the field, which includes maintaining awareness of new treatments, diagnostics, and research studies that could potentially benefit their patients and the ADRD field. Furthermore, new care models such as those proposed by Centers for Medicare and Medicaid Services (CMS)[31] demand additional resources that our current workforce is ill-equipped to support.

## Data Collection and Integration for Agentic AI

Data collection should first prioritize what clinicians can interpret and are currently using to diagnose neurodegenerative disease in both the clinic and research settings. In the following sections, we examine the key data types currently integral to ADRD diagnosis and how AI can enhance their collection and interpretation: the patient's history, physical examination, neuropsychological testing, neuroimaging, and biomarkers. Agentic AI systems will transform the collection of the patient's history, physical examination and neuropsychological testing, enabling consistency and standardization not currently possible when relying on providers with different training and practice patterns. These AI systems can then integrate this data alongside neuroimaging and biomarkers for interpretation.

### *Gathering the patient's history*

Recent advances in LLM capabilities have made comprehensive remote history collection with voice-based conversational agents feasible for the first time. This includes natural conversational interaction that can match or exceed clinician efficiency and expressions of empathy,[7] engage with persons living with dementia,[32] and adapt to various educational backgrounds and languages.[33]

Clinicians, faced with limited time, struggle to gather and document comprehensive patient histories from distressed patients or their caregivers, leading to redundant interviews across clinicians and potentially delaying diagnosis. Moreover, incomplete documentation restricts the valuable clinical data available for AI-driven decision support systems. Voice-based conversational agents powered by LLMs could transform history collection by enabling these interviews to be conducted over the phone outside time-limited clinic visits, increasing healthcare access for patients who are of lower socioeconomic status, have limited English proficiency, and/or have low digital literacy.[34–37] This can democratize the highly specialized interviews needed to differentiate between neurodegenerative conditions (e.g., syndromes most compatible with Alzheimer's disease, frontotemporal dementia, or Lewy Body disease) and could be integrated at various points along the patient's care pathway.

There are several considerations for the efficacy and safety of interactions between patients and conversational agents. Interviews will need to be semi-structured as LLMs could engage in irrelevant or dangerous topics. Continuous real-time monitoring systems are needed to analyze the conversation for content relevance and for dangerous content such as expressions of suicidal ideation, automatically triggering human intervention. A hybrid model that combines automated conversational agents with timely clinician oversight and decision-making is strongly advocated. Bias will also need to be identified and mitigated (e.g., these tools can have worse performance with non-English speaking people).[38]

Most clinicians would not realistically adopt a system requiring them to read a lengthy transcript, but LLM summarization capabilities enable workflow integration by converting these interviews into a *history of present illness* (HPI).[39] Reliable summarization is especially critical in cognitive impairment contexts, where accurate documentation of subtle symptom changes over time is essential for early diagnosis and management.

*Neuropsychological Testing*

When concerning symptoms are identified, a cognitive assessment with objective measures of cognitive impairment is required. Cognitive assessments should be scalable, without requiring highly trained staff, and designed to identify relative weaknesses of specific cognitive and behavioral domains to enable identification of patterns helpful with diagnosis, going beyond brief screening assessments. Traditional assessments are time-consuming, require specialized training, and often capture only summary scores rather than the detailed, domain-specific performance data valuable for AI interpretation.

Digital cognitive assessments will be necessary to enable a comprehensive AI system for ADRD, and they address the above limitations by capturing nuances in performance (trial-level data, response times, error patterns) unavailable with paper-based tests. Their ability to integrate directly into the clinician's workflow within the EHR has driven adoption, as shown with the TABCat-BHA.[40] The digital cognitive assessment landscape is advancing rapidly,[41,42] with clinically available tools continually improving in accuracy, accessibility, and ecological validity.[41,42] Virtual reality assessments represent one particularly promising direction,[43] with emerging LLM integration enabling virtual psychometrists that could fundamentally transform how cognitive evaluations are conducted.

Note that gathering the history and cognitive assessment ahead of the visit could free up time with the clinician, enabling a conversation designed to build rapport and shared decision making that a clinician would not otherwise have the time to do well.

*Physical Examination*

Physical examination continues to be important in obtaining a diagnosis and the motor findings associated with Lewy body disease, corticobasal syndrome, progressive supranuclear palsy, amyotrophic lateral sclerosis, cerebrovascular disease and many other conditions. In addition, the absence of motor findings in AD is also important. While an in-person physical exam is ideal, physical exam findings for these syndromes can often be uncovered through video visits which can increase accessibility for patients who cannot feasibly travel to see a specialist. Possibilities include video interpretation of speech and speech patterns, motor planning and execution, gait, balance, and others. New multimodal AI models have made significant progress in describing video content such as surgeries, making video data a promising method for AI decision support with further research.[44]

*Neuroimaging and Biofluid Biomarker Data*

Neuroimaging with MRI and Positron Emission Tomography (PET) is a diagnostic cornerstone, aiding in exclusion of treatable causes of dementia, assessing vascular etiologies, and observing atrophy patterns (MRI) or biomarker location (PET). MRI is becoming more accessible around the country with several creative solutions, so we do not see this as a notable bottleneck for the future of AI decision support.[45]

Biomarkers are now a valuable component of the diagnostic process. While amyloid PET and cerebrospinal fluid (CSF) biomarkers for Alzheimer's disease have not been easily scalable or accessible, multiple plasma-based biomarkers are becoming available in the clinic setting.[46–48] Assays of cerebrospinal fluid and more recently skin biopsies are promising for detection of alpha synuclein, which is a biomarker of Parkinsons' disease, Dementia with Lewy bodies, and multiple system atrophy.[49–51] The implementation and real world accuracy of these biomarkers is highly promising.[48] Experts in the field have emphasized that accurate interpretation of these biomarkers still requires clinical context,[52,53] which is a concept that can be extended to any data type. Therefore, an AI-system must integrate data across modalities with unique patient characteristics (e.g., age and kidney function significantly alter interpretation of new plasma AD biomarkers) to inform diagnosis and management.

*Multimodal Data Integration: Creating a Unified Clinical Picture*

Standardized data collection is the essential first step toward a timely diagnosis, enabling both personalized management plans and meaningful research contributions. Methods must be easy for patients and their caregivers as well as clinicians, as this can increase access to care and ensure data is from diverse patient populations and clinic settings. However, achieving global standardization at the point of care is extremely challenging in diverse, resource-constrained settings. Therefore, we advocate for an 'AI-first' data infrastructure approach.[54] Rather than imposing rigid collection protocols on frontline clinicians, this approach prioritizes backend curation and harmonization layers that can transform heterogeneous real-world clinical data into AI-ready formats. Standardized, high-quality, data conversion for each modality not only facilitates efficient storage and sharing but also underpins the continuous development of effective AI systems. The quality of an agentic AI system is dependent on the quality of the data that it learns from, making this phase the foundation of our roadmap.

Both existing and future datasets require proper curation according to FAIR principles (Findable, Accessible, Interoperable, and Reusable), including comprehensive metadata, clear governance frameworks, and quality control procedures.[55] National organizations such as the NIH, CMS, advocacy organizations, and other entities could potentially facilitate standards development (e.g., Digital Imaging and Communications in Medicine (DICOM) for neuroimaging) and monitoring, although there are many other governance models which have been written about elsewhere.[56]

The most challenging aspect of a scalable AI system in the US healthcare system will likely be multimodal data integration across diverse and complex EHR systems while maintaining strict HIPAA compliance and data protection standards. Practical pathways forward are forming with two decades of progress in interoperability standards and now greater momentum as AI has dramatically increased the value of data; however, implementation will be difficult across healthcare settings due to varying degrees of resource constraints, technical expertise, and competing institutional priorities.

The Substitutable Medical Apps, Reusable Technology (SMART) on Fast Healthcare Interoperability Resources (FHIR) framework enables applications to read and write standardized health data across vendors,[57,58] with major EHR systems like Epic and Oracle already supporting robust integration ecosystems.[59,60] Real-world examples demonstrate this capability. UCSF's BRIDGE dashboards integrate multimodal sources across any SMART on FHIR compatible EHR.[61] The Observational Medical Outcomes Partnership (OMOP) Common Data Model extension for imaging (MI-CDM) was used to convert 1 million DICOM series from an Alzheimer's neuroimaging cohort into OMOP tables alongside clinical EHR data.[62] Similar advances have integrated genomic, wearable, and digital cognitive assessment data into the EHR.[63,64] Research initiatives have created environments with standardized, multimodal data from multiple institutions, such as the Alzheimer's Disease Data Initiative (ADDI) and AD Workbench, a platform with the compute resources necessary for multi-institutional collaboration.[65] The Global Research and Imaging Platform (GRIP), a secure, cloud-based open-access ecosystem, extends this model by providing modular workflows, tools for integrating and analyzing complex multimodal datasets, and secure spaces for collaborative, community-driven development. In parallel, JupyterHealth, an open, modular software platform for health data ingestion and management, spans the full lifecycle of digital health workflows from data ingestion, data standardization and transformation, secure storage and authentication to data management and analysis within clinical workflows. Building on HL7 FHIR, Open mHealth and related interoperability standards, it shows how open, community-driven infrastructure can be adopted by diverse health systems, researchers, and clinicians.

LLMs are likely to accelerate interoperability further. Recent work demonstrated an LLM translating free-text EHR notes into FHIR with impressive accuracy, suggesting that LLMs may enable interoperability driven less by rigid schemas and more by flexible "semantic standards," dynamically reshaping data into whatever structure downstream systems require.[66]

Collectively, these advances demonstrate that multimodal EHR data integration is possible at institutional scales. However, current implementations remain limited to highly motivated, well-resourced institutions, with widespread adoption across diverse healthcare settings not yet realistic. Nevertheless, rapid policy and technical advances in interoperability create urgency to pilot AI systems integrating multimodal data now, positioning healthcare systems for broader deployment as these capabilities mature. Furthermore, to scale these solutions beyond research cohorts into routine care, we must address barriers that are not solely technical, but largely issues of governance. Divergent policies on data access and privacy across institutions create silos that hinder the training of robust AI models. An AI-ready framework requires not only technical interoperability (e.g., handling different data formats) but also semantic alignment and dynamic governance protocols that can manage consent and privacy across borders and institutions.

*Enhancing the Human Element and Laying the Foundation for AI Decision Support*

We envision that AI collecting data in a consistent and standardized manner (i.e., history, cognitive assessment, physical exam) will serve as a foundation for reducing errors and enhancing clinician-patient connection by shifting clinic visit time from data collection/documentation to more meaningful tasks: reviewing findings with patients, answering questions, and engaging in shared decision-making about care preferences and values. This data can be presented alongside already standardized neuroimaging and biomarkers in an explainable manner.

This infrastructure enables both immediate diagnostic capabilities and continuous learning, which will allow the AI system to adapt as diagnostic tools evolve (e.g., streamlining patient history as biomarkers become more accurate). Practically, implementation will likely begin with larger healthcare systems with the capacity for multimodal data sharing. Smaller systems will require demonstrated value before participating in voluntary data exchange programs such as the Trusted Exchange Framework and Common Agreement (TEFCA), a government initiative designed to facilitate nationwide health information exchange. With sustained effort and evidence of benefit, we can realize the full potential of an AI system serving as a decision support tool in ADRD care.

***AI Decision Support: Data Interpretation to Facilitate Diagnosis and Management***

High-quality data collection and standardization outlined above enables both the clinician and the AI system to efficiently access, integrate, and interpret multimodal patient data, avoiding the risk of false positives from relying on one test result. Our perspective on AI decision support centers on augmenting rather than replacing clinician capabilities by highlighting important diagnostic features from this complex multimodal data, explaining its reasoning transparently.[5,8,67] In this way, AI can be a collaborative partner, rather than a passive tool.

*AI Research in ADRD Decision Support*

Despite successes with disease classification tasks and their interpretability, traditional machine learning approaches (e.g., logistic regression, support vector machines, and random forests) face significant challenges with the multi-modal, heterogeneous datasets in ADRD research. These methods require extensive manual feature engineering to extract structured representations from diverse data types, ranging from MRIs and PET scans to tabular EHRs and unstructured clinical notes. This introduces potential biases and information loss as well as limits scalability and adaptability. More advanced computational models are needed capable of processing and integrating relational, unstructured, and multi-modal data to capture the nuanced, multifactorial nature of cognitive decline.

Deep learning architectures, particularly convolutional neural networks (CNNs), have enabled remarkable precision in identifying disease-specific features from neuroimaging[68], such as atrophy patterns or cortical thinning.[69] Recent multimodal AI models incorporating clinical, laboratory, and imaging data hold the most promise,[70–72] with recent successful multimodal models being able to differentiate neurodegenerative subtypes or stages of disease with high accuracy. Although commercial solutions exist for AI-based regional normalized brain volumetry, it remains unclear how much such methods are affected by image acquisition parameters and how such data are best incorporated into a diagnosis for an individual patient.[73–75]

Despite advances, deep learning models face significant implementation challenges. These include substantial data requirements, overfitting risk with small ADRD datasets, and sometimes limited interpretability.[76] This lack of transparency complicates clinical adoption by undermining provider trust, highlighting the need for robust validation protocols, diverse training datasets, and user interfaces that translate complex model inferences into clinically meaningful explanations.[77]

*Using Agentic AI for ADRD Decision Support*

Recent advancements in LLMs may introduce a paradigm shift in ADRD care and research, offering novel approaches to clinical data interpretation and diagnostic uncertainty reduction. Unlike traditional discriminative models constrained by rigid classification frameworks, LLMs process extensive natural language inputs, including clinical documentation and patient-generated speech, while generating probabilistic assessments from incomplete or ambiguous clinical data. They have demonstrated impressive medical knowledge that could serve as clinical decision support. For example, GPT-4 exceeded mean human performance on a neurology board-style examination,[78] and can provide neuropathologic differential diagnoses similar to experts.[79]

Successful integration of AI in ADRD necessitates a complementary approach utilizing both discriminative models for distinguishing well-defined populations and generative models for managing diagnostic uncertainty and hypothesis generation. This combination can be accomplished through an agentic AI system which enhances both the accuracy and generalizability of decision support tools.

Agentic AI systems can address a fundamental healthcare challenge: delivering the right information to the right patient at the right time. In practice, the clinician could review AI generated insights about the patient before the appointment, including a differential diagnosis with explanations and then provide recommendations for next steps in workup and management based on the latest evidence-based guidelines. This need is growing with the complexity of emerging therapies, particularly anti-amyloid therapies with numerous inclusion/exclusion criteria and safety requirements making patient selection difficult even for specialists. AI systems can systematically evaluate these complex criteria, reducing errors and helping identify suitable candidates for specific interventions earlier. Such systems could also suggest personalized interventions ranging from disease-modifying therapies to supportive care like psychotherapy, physical therapy, cognitive maintenance strategies including lifestyle modifications, as well as matching patients with clinical trials. During appointments, clinicians verify findings with patients, guide shared-decision making about next steps, and leverage AI to provide explanations optimized for that patient's level of education, age, and culture.[80]

A practical advantage of agentic AI in ADRD care is its ability to integrate multimodal clinical data into a single, clinician-interpretable narrative. These systems can combine text (clinic notes, radiology reports), structured data (labs, vitals, medications), and higher-dimensional signals (e.g. imaging, speech, actigraphy-based wearables) in two main ways. One approach is modality-to-language standardization where imaging features, test results, and other non-text input features are converted into textual or tabular summaries that can be consumed by LLM-based agents. Gallingani et al. (2025) developed an agentic AI system that simulates a multidisciplinary clinical team that achieved high accuracy in classifying pathology-confirmed primary progressive aphasia variants using multimodal inputs.[81] The multimodal includes clinical notes, neuropsychological and language assessments, and MRI-derived atrophy scores. The system preprocesses these data extracting features that are then converted into textual summaries and then uses specialized agents (for behavioral neurology, neuropsychology, and neuroimaging), whose analyses are synthesized and reviewed by a set of prompt-engineered specialized AI agents. Importantly, the output report is designed to be interpretable in familiar clinical terms that resemble how clinicians present cases, allowing clinicians to evaluate the evidence supporting the differential and the predicted underlying pathology. Microsoft's MAI-DxO (MAI Diagnostic Orchestrator) illustrates a related orchestration pattern: sequential diagnostic reasoning that surfaces clinician-readable intermediate steps and culminates in an auditable natural-language synthesis of what was asked, what was observed, and why a recommendation follows.[82] A second approach is to build on multimodal foundation models that reason jointly over vision and language, directly incorporating images or documents alongside text during dialogue and inference while maintaining a clinician-facing summary for review, as demonstrated by multimodal AMIE (Articulate Medical Intelligence Explorer), an agentic AI system developed by Google for telehealth-style consultations.[83] Together, these examples illustrate emerging agentic AI system architectures that support multimodal data integration and clinician-facing decision support.

*Economic Considerations of Clinical Integration*

Economic viability of healthcare AI systems remains uncertain, with substantial unknowns around return on investment (ROI), infrastructure costs, and reimbursement pathways. A recent systematic review demonstrated AI cost effectiveness across clinical domains from a healthcare system perspective through Quality of Life Years gained, Disability Adjusted Life Years averted, and diagnostic yield improvements, reducing unnecessary studies and treatments.[84] Other studies have found earlier diagnosis to be cost-effective for Alzheimer’s disease, particularly when accounting for caregiver economic burden and slowing progression with new disease modifying therapies.[85] However, there is still some uncertainty as many studies fail to fully account for infrastructure investments, integration expenses, and maintenance costs. Further, while the ROI using AI appears to be favorable for radiology,[86] the ROI for other applications are mainly estimates based on little data.

There are trends that offer cautious optimism. LLM costs are significantly declining, particularly inference and hardware costs, as well as improving energy efficiency.[87] Evidence suggests that value-based payment models could incentivize the adoption of AI, though currently reimbursement for AI is through per-use payments.[88] There is national government momentum in enabling new reimbursement pathways as demonstrated by a recent bipartisan bill introduced in April of 2025 (The Health Tech Investment Act, or S. 1399), which proposes systematic Medicare coverage for newly FDA authorized AI enabled medical devices to promote AI innovation. Healthcare systems have already shown willingness to invest in AI without clear reimbursement or concrete dollar ROI as demonstrated by the rapid adoption of ambient scribes, with evidence of decreasing clinician burnout.[89] While a shared AI system could provide economies of scale among multiple healthcare systems, government funding would likely be needed to incentivize the upfront costs, and more reliable reimbursement may be needed to cover ongoing maintenance costs.

Electronic consultations (e-consults) offer the most practical near-term implementation pathway, which are asynchronous, provider-to-provider communications that allow clinicians such as PCPs to seek specialist input through a shared electronic health record or web-based platform, often resulting in fewer in person specialist visits.[90] In our experience, e-consults for a cognitive neurologist typically takes less than 10 minutes to complete. However, due to the limited data available with these consults many still require a long wait for a 60+ minute in person visit. The AI system proposed here could provide the data and decision support necessary for specialists to definitively address more cases in a fraction of the time required for a new patient visit, reducing wait times and patient travel burden especially for rural populations. Furthermore, specialists performing these reviews would simultaneously validate AI system performance, creating a sustainable monitoring mechanism without additional time investment. Demonstrating economic sustainability of these systems will be essential before widespread implementation.

**Validation and Monitoring**

Validation requires comparison with appropriately defined ‘gold standards” in a generalizable clinical population, with sufficient statistical power to demonstrate efficacy on a patient by patient (not simply group wise) basis. The vast majority of generative AI diagnostic studies rely on simulated rather than real-world data, highlighting the need for more systematic clinical validation.[91] Prior to real world implementation, the possibility of misdiagnosis or adverse events must be virtually zero. Most clinical research with the gold standard neuropathologic verification is performed on well-educated relatively affluent patients with health insurance, often in academic settings. However, in the real world, patients have less education and insurance, greater comorbidities, and may be less fluent in English. Therefore, we propose models should also be developed using diverse specialist-confirmed cases in addition to neuropathology. This approach acknowledges both the gold standard status of neuropathology and the practical need for representative validation.

However, specialist diagnoses contain uncertainty and there is variability even within a single practitioner, with greater variability across different clinicians and institutions. Careful case selection is required for reliable validation datasets in the absence of neuropathological confirmation. For example, prioritizing cases confirmed through highly accurate biomarkers with matching syndromes (e.g., Alzheimer’s disease biomarkers), clearly meeting criteria known to be highly specific for a certain pathology (e.g., Progressive Supranuclear Palsy Richardson syndrome), multiple specialist agreement, or longitudinal confirmation of diagnosis through disease progression patterns. This step is crucial to reduce the risk of contaminated model updates.

Ongoing monitoring is essential to prevent model degradation—the unintended decline in model performance, accuracy, or safety from improper updates, contaminated training data, or inconsistent validation practices. Agreed upon benchmarks must ensure models are fair, appropriate, valid, effective, and safe (FAVES).[92] Recent initiatives like OpenAI's HealthBench and Stanford's MedArena demonstrate promising approaches for systematic medical AI evaluation across diverse clinical scenarios. These benchmarking frameworks could be adapted for neurodegenerative disease-specific applications, continuously tested on the unique diagnostic challenges these conditions present.

Specialists are best positioned to identify model errors. Much of this monitoring can occur naturally through AI system use. Specialists could use this system to improve their own efficiency and quality in patient visits and e-consults, positioning them to detect errors in routine practice. While specialist time is limited, their investment in developing AI systems scales their expertise enabling PCPs and general neurologists to definitively manage more cases, reducing unnecessary referrals and increasing their capacity. This represents an evolution in specialist roles, where expertise is leveraged through optimizing AI systems rather than confined to individual patient encounters. If cost-effectiveness evidence builds, reimbursement mechanisms could justify dedicated monitoring activities such as batch evaluation of flagged cases. Beyond technical performance, monitoring must evaluate physician-AI interaction patterns, as clinicians often over rely on or underutilize AI capabilities,[93,94] informing both system design and clinician training. In real-world settings, it's equally important to monitor how AI use affects clinician efficiency, productivity, and job satisfaction.

To ensure coherence across the AI lifecycle, the oversight body governing standards would ideally also oversee validation, monitoring, and continuous learning processes discussed in the next section. While validation and monitoring focuses on quality control and preventing harm, and continuous learning aims to identify opportunities for improvement, these functions are inseparable: continuous learning without rigorous validation risks model degradation from contaminated or unverified updates, while validation without continuous improvement fails to leverage clinical experience. The careful case selection criteria described above apply equally to learning datasets. A unified governance approach ensures consistent standards, maintains institutional knowledge about AI system behavior, and aligns system evolution with healthcare priorities rather than commercial interests.

### Continuous Learning Guided by the ADRD Community

While validation and monitoring prevent harm, continuous learning ensures this agentic AI system advances with emerging knowledge and clinical experience. This requires staying up to date with the rapidly evolving medical literature as well as learning from carefully validated patient cases in a supervised manner, guided collaboratively by the healthcare community.

Clinician learning through patient cases is currently limited as they do not often see what happens to a patient if that patient does not follow up with them over time. Furthermore, what one clinician learns is difficult to transmit to others; however, an agentic AI system can learn from and provide benefit to every clinician using it (see Figure 1). Built in feedback loops enable the AI system to consistently revisit each patient's chart or other databases to check for updates, such as an updated diagnosis or treatment response. Clinicians could receive these updates as well, learning alongside the AI system and taking action when needed. The careful case selection principles from the Validation section apply to continuous learning: only high-quality, validated cases from diverse clinical settings should influence model evolution, with ongoing oversight preventing contaminated updates.

**Figure 1 – Continuous Improvement from Specialist Use**

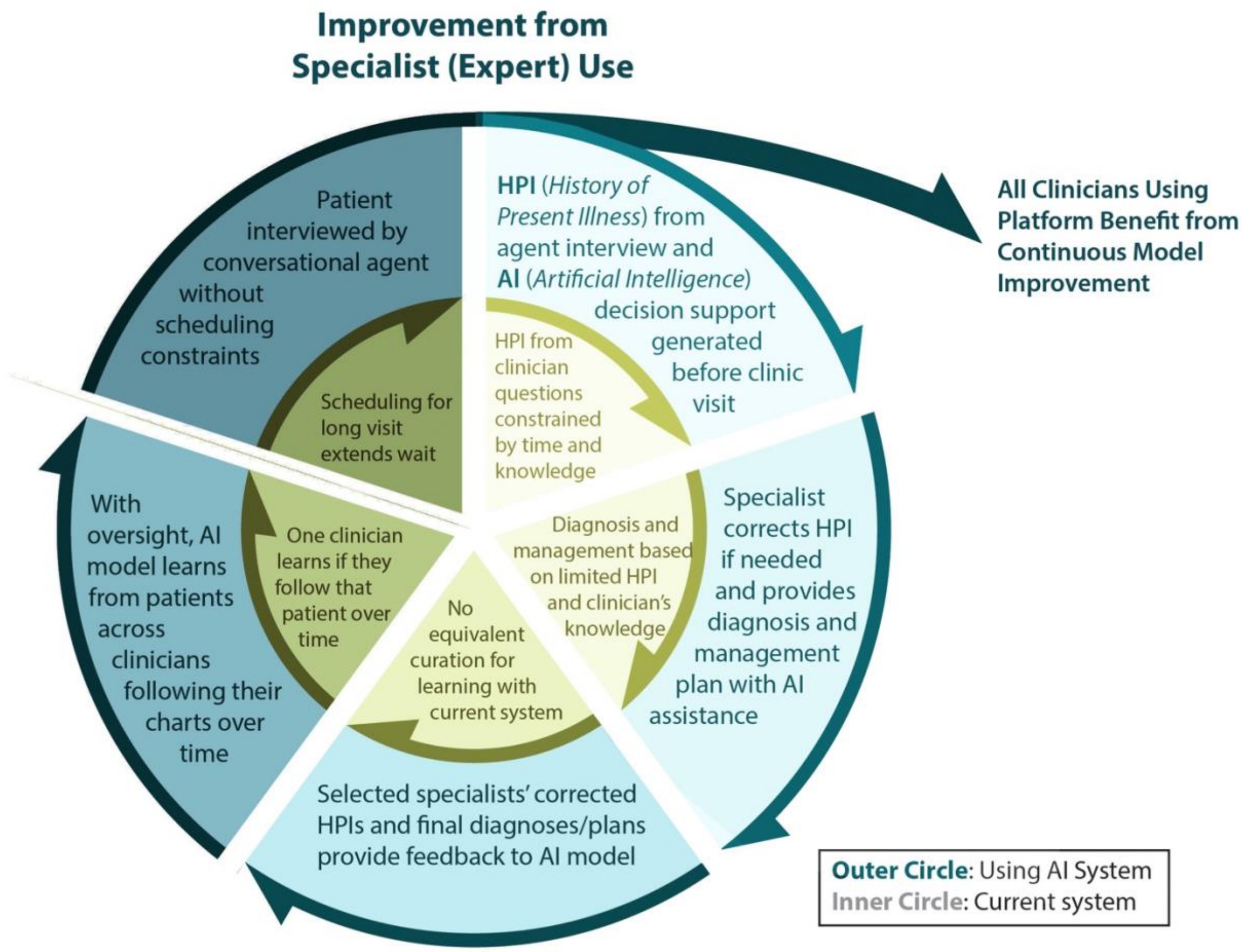

Transparency in AI development can be accomplished through using open-source/open-weights models with robust governance. When proprietary models are updated to optimize general use cases like writing or coding, performance on specific use cases such as healthcare can degrade unpredictably, requiring costly and frequent revalidation cycles with many models sunsetting every 1-2 years. Open-source models offer full control over updates, enable complete fine-tuning, and allow mechanistic interpretability analysis (i.e., understanding how an AI model is making decisions). LLMs of smaller size (8B-70B parameters) can be trained to match or surpass the performance of large proprietary models for more specific, narrower tasks including the task of answering medical questions,[95–97] making subspecialty-focused models feasible. Although the computational overhead of these models once made routine clinical deployment impractical, recent parameter-efficient LLM compression methods (for example, 4-bit QLoRA[98] or aggressive edge-device quantization) now enable low-latency, on-device inference suitable for real-world clinical workflows. Figure 2 provides an overview on how this AI system could evolve over time.

**<u>Figure 2 – Evolution of Agentic AI Systems</u>**

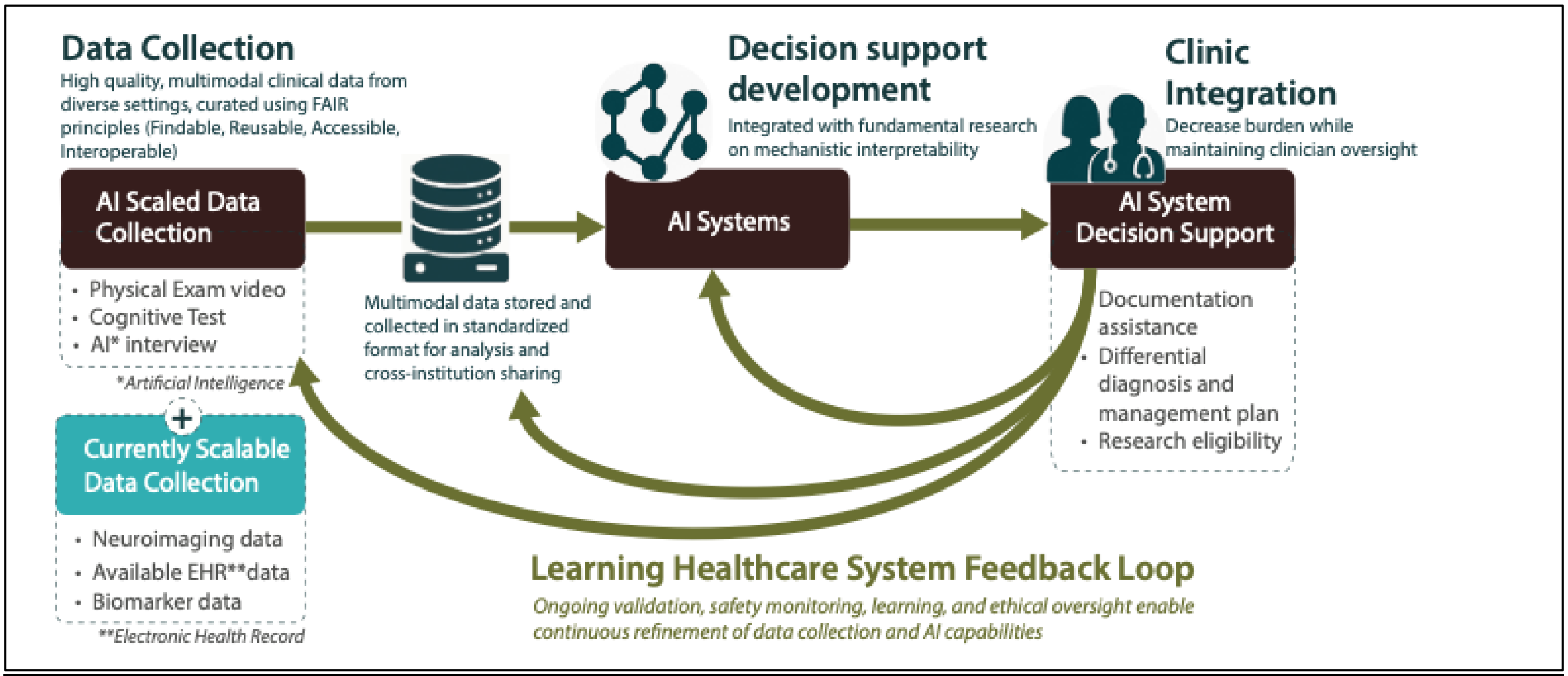

### *Ethical Use of AI*

Ethical challenges around the use of AI in ADRD have been discussed elsewhere,[99] and therefore, a brief overview is provided here with updates that apply across healthcare. ADRD populations face unique ethical considerations beyond general AI concerns. As a vulnerable population with declining capacity, autonomy changes over time, requiring informed consent processes that evolve dynamically to involve both patients and caregivers appropriately. This shifting autonomy creates inherent tension between safety monitoring and preserving patient independence and dignity. Design considerations risk amplifying disparities if cognitive decline disproportionately impairs patients' ability to use AI tools, while privacy concerns intensify given the sensitive cognitive and behavioral data collected from individuals who may lack capacity to understand data usage.

Many ethical considerations have intensified around the use of AI in healthcare, particularly since the widespread enthusiasm with generative AI since November 2022. In response, numerous initiatives have developed frameworks and guidelines to ensure AI applications are fair, appropriate, valid, effective, and safe (FAVES), which will at least in part require transparency with technical performance as well as organizational capacity to manage risks associated with deploying this technology.[92] Fairness has been of particular concern as AI has shown numerous biases,[100,101] emphasizing the need for training data from diverse groups as well as ongoing monitoring to identify and quickly mitigate biases based upon the analysis of only socioeconomically privileged populations. The advancement in AI capabilities will bring new challenges with patient privacy and informed consent.[102,103] When using clinical notes for example, removing PHI may be insufficient to protect privacy,[104] potentially requiring more stringent data protection and sharing guidelines. Regardless of the quality of the training data, AI tools are not static, in contrast to some pharmaceuticals whose mechanisms of action may be unclear but are safely used with patients. Therefore, explainability and interpretability[105–107] are important components for a human in the loop to potentially identify signs of model deterioration or drift at the point of care. Ethical considerations traditionally designed to protect patients will likely need to be extended to healthcare workers as well with the rapid evolution of generative AI models.[108]

For example, generative AI's ability to mimic providers' appearance, voice, and empathetic conversation will likely be of interest to healthcare employers with an overwhelming patient demand. This dynamic landscape has prompted calls for a network of Health AI Ethics Centers to adapt and respond to these ongoing changes to keep pace in a manner that is often difficult for a central governmental agency.[108]

We believe patient harm represents the greatest risk of AI implementation in healthcare. This may occur through privacy breaches, adverse psychological effects from AI interactions (including anxiety, depression, or suicidal ideation), misdiagnosis, inappropriate treatment recommendations, medication errors, and more. Given the enthusiasm for medical AI and the economic incentives driving rapid adoption, we are concerned these risks are underappreciated. The possibility of widespread harm is not insignificant, particularly with an agentic AI system that is given too much autonomy. Legal liability remains unclear—who bears responsibility if patients suffer or die due to AI errors? Therefore, we emphasize that implementing redundant monitoring systems and establishing clear reporting mechanisms is critical to ensure prompt remediation when adverse events inevitably occur.

The current fragmented governance landscape in healthcare AI is still evolving.[109] While there has been a significant increase in FDA approved AI-enabled medical devices, the FDA has limited regulatory authority over many healthcare AI applications (e.g., they do not regulate applications that only provide recommendations or information that clinicians do not primarily rely on). Other federal agencies such as CMS have regulatory authority over AI implementation limited to entities participating in their programs. In the absence of comprehensive federal regulation, governance is emerging through state legislation, nonprofit guidance (e.g., the Joint Commission and Coalition for Health AI providing national guidance for responsible AI use), and private contracts between developers and healthcare organizations.[109] AI integration in diagnostic workflows will require greater involvement from policymakers and government agencies alongside non-governmental entities. Given this fragmented landscape, the unified oversight approach we propose becomes even more essential: integrating data collection standards, validation protocols, monitoring systems, and continuous learning under collaborative governance. This model demonstrates how healthcare specialty communities can establish rigorous governance frameworks that complement and collaborate with evolving governmental regulation, addressing current gaps while maintaining alignment with evolving regulatory requirements.

**Conclusion**

The integration of agentic AI systems into clinical care offers transformative potential to address many of the most pressing challenges in this unprecedented time for the field of ADRD. In an era with increasing time constraints and an overwhelming volume of medical knowledge, agentic AI systems can support clinicians by enhancing data collection, improving diagnostic accuracy, and personalizing management, ultimately enabling more timely and high-quality patient care. Successful implementation of these technologies will require a rigorous framework that prioritizes inclusivity, transparency, and ongoing monitoring to ensure ethical and effective use. Collaboration among clinicians, researchers, policymakers, governmental agencies, industry, and patients will be critical to establish standards for data collection, model development, and real-world application, ultimately fostering a continuously learning healthcare system.

While this roadmap focuses on ADRD care, its principles apply broadly across neurology and other specialties facing similar challenges. With thoughtful development, agentic AI systems can not only mitigate provider shortages and diagnostic delays but also redefine care delivery, creating a future where technology and human expertise work in harmony to improve patient outcomes and advance medical knowledge.

## Author Contributions

All authors meet the ICMJE requirements for authorship through 1) substantial contributions to the conception of the work, 2) reviewing the work critically for important intellectual content, 3) final approval of the version to be published, and 4) agreement to be accountable for all aspects of the work in ensuring that questions related to the accuracy or integrity of any part of the work are appropriately investigated and resolved.

## Competing Interests Statement

Andreas Rauschecker reports receiving research funding from GE Healthcare.

## REFERENCES:

1. Majersik JJ, Ahmed A, Chen IHA, et al. A Shortage of Neurologists – We Must Act Now: A Report From the AAN 2019 Transforming Leaders Program. *Neurology*. 2021;96(24):1122-1134. doi:10.1212/WNL.0000000000012111

2. Nourazari S, Hoch DB, Capawanna S, Sipahi R, Benneyan JC. Can improved specialty access moderate emergency department overuse? *Neurol Clin Pract*. 2016;6(6):498-505. doi:10.1212/CPJ.0000000000000295

3. Pizer SD, Prentice JC. What Are the Consequences of Waiting for Health Care in the Veteran Population? *J Gen Intern Med*. 2011;26(Suppl 2):676-682. doi:10.1007/s11606-011-1819-1

4. Singhal K, Azizi S, Tu T, et al. Large language models encode clinical knowledge. *Nature*. 2023;620(7972):172-180. doi:10.1038/s41586-023-06291-2

5. Singhal K, Tu T, Gottweis J, et al. Toward expert-level medical question answering with large language models. *Nat Med*. Published online January 8, 2025:1-8. doi:10.1038/s41591-024-03423-7

6. Tu T, Azizi S, Driess D, et al. Towards Generalist Biomedical AI. *NEJM AI*. 2024;1(3):AIoa2300138. doi:10.1056/AIoa2300138

7. Tu T, Schaekermann M, Palepu A, et al. Towards conversational diagnostic artificial intelligence. *Nature*. Published online April 9, 2025:1-9. doi:10.1038/s41586-025-08866-7

8. Liu X, Liu H, Yang G, et al. A generalist medical language model for disease diagnosis assistance. *Nat Med*. Published online January 8, 2025:1-11. doi:10.1038/s41591-024-03416-6

9. Gupta R, Ghodsi MZ Omar Khattab, Lingjiao Chen, Jared Quincy Davis, Heather Miller, Chris Potts, James Zou, Michael Carbin, Jonathan Frankle, Naveen Rao, Ali. The Shift from Models to Compound AI Systems. The Berkeley Artificial Intelligence Research Blog. Accessed May 24, 2025. http://bair.berkeley.edu/blog/2024/02/18/compound-ai-systems/

10. Acharya DB, Kuppan K, Divya B. Agentic AI: Autonomous Intelligence for Complex Goals—A Comprehensive Survey. *IEEE Access*. 2025;13:18912-18936. doi:10.1109/ACCESS.2025.3532853

11. For trustworthy AI, keep the human in the loop. *Nat Med*. 2025;31(10):3207-3207. doi:10.1038/s41591-025-04033-7

12. Dell'Acqua F, McFowland E, Mollick ER, et al. Navigating the Jagged Technological Frontier: Field Experimental Evidence of the Effects of AI on Knowledge Worker Productivity and Quality. *SSRN Electron J*. Published online 2023. doi:10.2139/ssrn.4573321

13. Gargari OK, Habibi G. Enhancing medical AI with retrieval-augmented generation: A mini narrative review. *Digit Health*. 2025;11:20552076251337177. doi:10.1177/20552076251337177

14. Shi H, Xu Z, Wang H, et al. Continual Learning of Large Language Models: A Comprehensive Survey. *ACM Comput Surv*. 2025;58(5):120:1-120:42. doi:10.1145/3735633

15. Liu NF, Lin K, Hewitt J, et al. Lost in the Middle: How Language Models Use Long Contexts. *Trans Assoc Comput Linguist*. 2024;12:157-173. doi:10.1162/tacl_a_00638

16. He J, Zhang B, Rouhizadeh H, et al. Retrieval-Augmented Generation in Biomedicine: A Survey of Technologies, Datasets, and Clinical Applications. *arXiv*. Preprint posted online November 23, 2025. doi:10.48550/arXiv.2505.01146

17. Agrawal R, Gupta T, Gupta S, Chauhan S, Patel P, Hamdare S. Fostering trust and interpretability: integrating explainable AI (XAI) with machine learning for enhanced disease prediction and decision transparency. *Diagn Pathol*. 2025;20:105. doi:10.1186/s13000-025-01686-3

18. Hebert LE, Weuve J, Scherr PA, Evans DA. Alzheimer disease in the United States (2010–2050) estimated using the 2010 census. *Neurology*. 2013;80(19):1778-1783. doi:10.1212/WNL.0b013e31828726f5

19. Global, regional, and national burden of neurological disorders, 1990–2016: a systematic analysis for the Global Burden of Disease Study 2016. *Lancet Neurol*. 2019;18(5):459-480. doi:10.1016/S1474-4422(18)30499-X

20. Fang M, Hu J, Weiss J, et al. Lifetime risk and projected burden of dementia. *Nat Med*. Published online January 13, 2025:1-5. doi:10.1038/s41591-024-03340-9

21. Bradford A, Kunik ME, Schulz P, Williams SP, Singh H. Missed and Delayed Diagnosis of Dementia in Primary Care: Prevalence and Contributing Factors. *Alzheimer Dis Assoc Disord*. 2009;23(4):306-314. doi:10.1097/WAD.0b013e3181a6bebc

22. Kotagal V, Langa KM, Plassman BL, et al. Factors associated with cognitive evaluations in the United States. *Neurology*. 2015;84(1):64-71. doi:10.1212/WNL.0000000000001096

23. Babulal GM, Quiroz YT, Albensi BC, et al. Perspectives on ethnic and racial disparities in Alzheimer's disease and related dementias: Update and areas of immediate need. *Alzheimers Dement J Alzheimers Assoc*. 2019;15(2):292-312. doi:10.1016/j.jalz.2018.09.009

24. Boustani M, Callahan CM, Unverzagt FW, et al. Implementing a Screening and Diagnosis Program for Dementia in Primary Care. *J Gen Intern Med*. 2005;20(7):572-577. doi:10.1111/j.1525-1497.2005.0126.x

25. Bernstein A, Rogers KM, Possin KL, et al. Primary Care Provider Attitudes and Practices Evaluating and Managing Patients with Neurocognitive Disorders. *J Gen Intern Med*. 2019;34(9):1691-1692. doi:10.1007/s11606-019-05013-7

26. Satiani A, Niedermier J, Satiani B, Svendsen DP. Projected Workforce of Psychiatrists in the United States: A Population Analysis. *Psychiatr Serv*. 2018;69(6):710-713. doi:10.1176/appi.ps.201700344

27. The American Geriatric Society. *PROJECTED FUTURE NEED FOR GERIATRICIANS*.; 2016:1-2. Accessed September 7, 2023. https://www.americangeriatrics.org/sites/default/files/inline-files/Projected-Future-Need-for-Geriatricians.pdf

28. *Modeling Early Detection and Geographic Variation in Health System Capacity for Alzheimer's Disease–Modifying Therapies*. RAND Corporation; 2024. doi:10.7249/RRA2643-1

29. van Dyck CH, Swanson CJ, Aisen P, et al. Lecanemab in Early Alzheimer's Disease. *N Engl J Med*. 2023;388(1):9-21. doi:10.1056/NEJMoa2212948

30. Sims JR, Zimmer JA, Evans CD, et al. Donanemab in Early Symptomatic Alzheimer Disease: The TRAILBLAZER-ALZ 2 Randomized Clinical Trial. *JAMA*. 2023;330(6):512-527. doi:10.1001/jama.2023.13239

31. Guiding an Improved Dementia Experience (GUIDE) Model | CMS. Accessed April 18, 2024. https://www.cms.gov/priorities/innovation/innovation-models/guide

32. Sheehy L, Bouchard S, Kakkar A, El Hakim R, Lhoest J, Frank A. Development and Initial Testing of an Artificial Intelligence-Based Virtual Reality Companion for People Living with Dementia in Long-Term Care. *J Clin Med*. 2024;13(18):5574. doi:10.3390/jcm13185574

33. Ray M, Kats DJ, Moorkens J, et al. Evaluating a Large Language Model in Translating Patient Instructions to Spanish Using a Standardized Framework. *JAMA Pediatr*. 2025;179(9):1026-1033. doi:10.1001/jamapediatrics.2025.1729

34. Tong L, George B, Crotty BH, et al. Telemedicine and health disparities: Association between patient characteristics and telemedicine, in-person, telephone and message-based care during the COVID-19 pandemic. *Ipem-Transl*. 2022;3:100010. doi:10.1016/j.ipemt.2022.100010

35. Nota SPFT, Strooker JA, Ring D. Differences in response rates between mail, e-mail, and telephone follow-up in hand surgery research. *Hand N Y N*. 2014;9(4):504-510. doi:10.1007/s11552-014-9618-x

36. Pagán VM, McClung KS, Peden CJ. An Observational Study of Disparities in Telemedicine Utilization in Primary Care Patients Before and During the COVID-19 Pandemic. *Telemed E-Health*. 2022;28(8):1117-1125. doi:10.1089/tmj.2021.0412

37. Hinton L, Franz CE, Reddy G, Flores Y, Kravitz RL, Barker JC. Practice constraints, behavioral problems, and dementia care: primary care physicians' perspectives. *J Gen Intern Med*. 2007;22(11):1487-1492. doi:10.1007/s11606-007-0317-y

38. Qu Y, Wang J. Performance and biases of Large Language Models in public opinion simulation. *Humanit Soc Sci Commun*. 2024;11(1):1-13. doi:10.1057/s41599-024-03609-x

39. Van Veen D, Van Uden C, Blankemeier L, et al. Adapted large language models can outperform medical experts in clinical text summarization. *Nat Med*. 2024;30(4):1134-1142. doi:10.1038/s41591-024-02855-5

40. Breithaupt AG, Sideman AB, Goode C, et al. Enhancing early detection of cognitive impairment in primary care with the TabCAT-BHA. *Alzheimers Dement*. 2025;21(7):e70437. doi:10.1002/alz.70437

41. Chen L, Zhen W, Peng D. Research on digital tool in cognitive assessment: a bibliometric analysis. *Front Psychiatry*. 2023;14. doi:10.3389/fpsyt.2023.1227261

42. Belleville S, LaPlume AA, Purkart R. Web-based cognitive assessment in older adults: Where do we stand? *Curr Opin Neurol*. 2023;36(5):491-497. doi:10.1097/WCO.0000000000001192

43. Mancuso V, Sarcinella ED, Bruni F, et al. Systematic review of memory assessment in virtual reality: evaluating convergent and divergent validity with traditional neuropsychological measures. *Front Hum Neurosci*. 2024;18. doi:10.3389/fnhum.2024.1380575

44. Saab K, Tu T, Weng WH, et al. Capabilities of Gemini Models in Medicine. *arXiv*. Preprint posted online May 1, 2024. Accessed September 26, 2024. http://arxiv.org/abs/2404.18416

45. Zhang K, Ganjizadeh A, Vahdati S, et al. Recent Advances in Compact Portable Platforms and Gradient Hardware for Brain MRI. *Radiology*. 2025;315(3):e241904. doi:10.1148/radiol.241904

46. Ashton NJ, Brum WS, Di Molfetta G, et al. Diagnostic Accuracy of a Plasma Phosphorylated Tau 217 Immunoassay for Alzheimer Disease Pathology. *JAMA Neurol*. 2024;81(3):255-263. doi:10.1001/jamaneurol.2023.5319

47. Janelidze S, Barthélemy NR, Salvadó G, et al. Plasma Phosphorylated Tau 217 and Aβ42/40 to Predict Early Brain Aβ Accumulation in People Without Cognitive Impairment. *JAMA Neurol*. 2024;81(9):947-957. doi:10.1001/jamaneurol.2024.2619

48. Palmqvist S, Tideman P, Mattsson-Carlgren N, et al. Blood Biomarkers to Detect Alzheimer Disease in Primary Care and Secondary Care. *JAMA*. Published online July 28, 2024. doi:10.1001/jama.2024.13855

49. Arnold MR, Coughlin DG, Brumbach BH, et al. α-Synuclein Seed Amplification in CSF and Brain from Patients with Different Brain Distributions of Pathological α-Synuclein in the Context of Co-Pathology and Non-LBD Diagnoses. *Ann Neurol*. 2022;92(4):650-662. doi:10.1002/ana.26453

50. Gibbons CH, Levine T, Adler C, et al. Skin Biopsy Detection of Phosphorylated α-Synuclein in Patients With Synucleinopathies. *JAMA*. 2024;331(15):1298-1306. doi:10.1001/jama.2024.0792

51. Samudra N, Fischer DL, Lenio S, et al. Clinicopathological correlation of cerebrospinal fluid alpha-synuclein seed amplification assay in a behavioral neurology autopsy cohort. *Alzheimers Dement*. 2024;20(5):3334-3341. doi:10.1002/alz.13799

52. Bouwman FH, Frisoni GB, Johnson SC, et al. Clinical application of CSF biomarkers for Alzheimer's disease: From rationale to ratios. *Alzheimers Dement Diagn Assess Dis Monit*. 2022;14(1):e12314. doi:10.1002/dad2.12314

53. Dubois B, Villain N, Schneider L, et al. Alzheimer Disease as a Clinical-Biological Construct—An International Working Group Recommendation. *JAMA Neurol*. Published online November 1, 2024. doi:10.1001/jamaneurol.2024.3770

54. Jasodanand VH, Bellitti M, Kolachalama VB. An AI-first framework for multimodal data in Alzheimer's disease and related dementias. *Alzheimers Dement*. 2025;21(9):e70719. doi:10.1002/alz.70719

55. Wilkinson MD, Dumontier M, Aalbersberg IjJ, et al. The FAIR Guiding Principles for scientific data management and stewardship. *Sci Data*. 2016;3(1):160018. doi:10.1038/sdata.2016.18

56. Shah NH, Halamka JD, Saria S, et al. A Nationwide Network of Health AI Assurance Laboratories. *JAMA*. 2024;331(3):245-249. doi:10.1001/jama.2023.26930

57. HL7 International. FHIR R5: Fast Healthcare Interoperability Resources. 2023. https://www.hl7.org/fhir/index.html

58. https://smarthealthit.org/. SMART Health IT. Accessed October 25, 2025. https://smarthealthit.org/

59. Oracle Health. Oracle Health Millennium Platform APIs. 2025. https://docs.oracle.com/en/industries/health/millennium-platform-apis/index.html

60. Epic Systems. Epic's home for data sharing and developer tools. 2025. https://open.epic.com/

61. Brown EG, Schleimer E, Bledsoe IO, et al. Enhancing clinical information display to improve patient encounters: Human-centered design and evaluation of the Parkinson Disease-BRIDGE platform. *JMIR Hum Factors*. 2022;9(2):e33967. doi:10.2196/33967

62. Park WY, Sippel Schmidt T, Salvador G, et al. Breaking data silos: incorporating the DICOM imaging standard into the OMOP CDM to enable multimodal research. *J Am Med Inform Assoc*. 2025;32(10):1533-1541. doi:10.1093/jamia/ocaf091

63. Lau-Min KS, McKenna D, Asher SB, et al. Impact of integrating genomic data into the electronic health record on genetics care delivery. *Genet Med*. 2022;24(11):2338-2350. doi:10.1016/j.gim.2022.08.009

64. Hoffman SL, Schmiedmayer P, Gala AS, et al. Comprehensive real time remote monitoring for Parkinson's disease using Quantitative DigitoGraphy. *NPJ Park Dis*. 2024;10(1):137. doi:10.1038/s41531-024-00751-w

65. McHugh CP, Clement MHS, Phatak M. AD Workbench: Transforming Alzheimer's research with secure, global, and collaborative data sharing and analysis. *Alzheimers Dement*. 2025;21(5):e70278. doi:10.1002/alz.70278

66. Brat GA, Mandel JC, McDermott MBA. Do We Need Data Standards in the Era of Large Language Models? *NEJM AI*. 2024;1(8):AIe2400548. doi:10.1056/AIe2400548

67. Savage T, Nayak A, Gallo R, Rangan E, Chen JH. Diagnostic reasoning prompts reveal the potential for large language model interpretability in medicine. *NPJ Digit Med*. 2024;7(1):20. doi:10.1038/s41746-024-01010-1

68. Esteva A, Robicquet A, Ramsundar B, et al. A guide to deep learning in healthcare. *Nat Med*. 2019;25(1):24-29. doi:10.1038/s41591-018-0316-z

69. Bron EE, Smits M, van der Flier WM, et al. Standardized evaluation of algorithms for computer-aided diagnosis of dementia based on structural MRI: The CADDementia challenge. *NeuroImage*. 2015;111:562-579. doi:10.1016/j.neuroimage.2015.01.048

70. Xue C, Kowshik SS, Lteif D, et al. AI-based differential diagnosis of dementia etiologies on multimodal data. *Nat Med*. 2024;30(10):2977-2989. doi:10.1038/s41591-024-03118-z

71. Rauschecker AM, Rudie JD, Xie L, et al. Artificial Intelligence System Approaching Neuroradiologist-level Differential Diagnosis Accuracy at Brain MRI. *Radiology*. 2020;295(3):626-637. doi:10.1148/radiol.2020190283

72. Leming M, Im H. Differential dementia detection from multimodal brain images in a real-world dataset. *Alzheimers Dement J Alzheimers Assoc*. 2025;21(7):e70362. doi:10.1002/alz.70362

73. Tanabe J, Lim MF, Dash S, et al. Automated Volumetric Software in Dementia: Help or Hindrance to the Neuroradiologist? *AJNR Am J Neuroradiol*. 2024;45(11):1737-1744. doi:10.3174/ajnr.A8406

74. Haller S, Falkovskiy P, Meuli R, et al. Basic MR sequence parameters systematically bias automated brain volume estimation. *Neuroradiology*. 2016;58(11):1153-1160. doi:10.1007/s00234-016-1737-3

75. Wittens MMJ, Sima DM, Houbrechts R, et al. Diagnostic Performance of Automated MRI Volumetry by icobrain dm for Alzheimer's Disease in a Clinical Setting: A REMEMBER Study. *J Alzheimers Dis JAD*. 2021;83(2):623-639. doi:10.3233/JAD-210450

76. Yu KH, Beam AL, Kohane IS. Artificial intelligence in healthcare. *Nat Biomed Eng*. 2018;2(10):719-731. doi:10.1038/s41551-018-0305-z

77. Aristidou A, Jena R, Topol EJ. Bridging the chasm between AI and clinical implementation. *The Lancet*. 2022;399(10325):620. doi:10.1016/S0140-6736(22)00235-5

78. Schubert MC, Wick W, Venkataramani V. Performance of Large Language Models on a Neurology Board–Style Examination. *JAMA Netw Open*. 2023;6(12):e2346721. doi:10.1001/jamanetworkopen.2023.46721

79. Koga S, Martin NB, Dickson DW. Evaluating the performance of large language models: ChatGPT and Google Bard in generating differential diagnoses in clinicopathological conferences of neurodegenerative disorders. *Brain Pathol*. 2023;34(3):e13207. doi:10.1111/bpa.13207

80. Ayers JW, Poliak A, Dredze M, et al. Comparing Physician and Artificial Intelligence Chatbot Responses to Patient Questions Posted to a Public Social Media Forum. *JAMA Intern Med*. 2023;183(6):589-596. doi:10.1001/jamainternmed.2023.1838

81. Gallingani C, Miller ZA, Mandelli ML, et al. Agentic Generative Artificial Intelligence System for Classification of Pathology-Confirmed Primary Progressive Aphasia Variants. *medRxiv*. Preprint posted online October 29, 2025:2025.10.28.25338977. doi:10.1101/2025.10.28.25338977

82. Nori H, Daswani M, Kelly C, et al. Sequential Diagnosis with Language Models. *arXiv*. Preprint posted online July 2, 2025. doi:10.48550/arXiv.2506.22405

83. Saab K, Freyberg J, Park C, et al. Advancing Conversational Diagnostic AI with Multimodal Reasoning. *arXiv*. Preprint posted online May 6, 2025. doi:10.48550/arXiv.2505.04653

84. El Arab RA, Al Moosa OA. Systematic review of cost effectiveness and budget impact of artificial intelligence in healthcare. *Npj Digit Med*. 2025;8(1):548. doi:10.1038/s41746-025-01722-y

85. Aye S, Handels R, Winblad B, Jönsson L. Optimising Alzheimer's Disease Diagnosis and Treatment: Assessing Cost-Utility of Integrating Blood Biomarkers in Clinical Practice for Disease-Modifying Treatment. *J Prev Alzheimers Dis*. 2024;11(4):928-942. doi:10.14283/jpad.2024.67

86. Bharadwaj P, Nicola L, Breau-Brunel M, et al. Unlocking the Value: Quantifying the Return on Investment of Hospital Artificial Intelligence. *J Am Coll Radiol JACR*. 2024;21(10):1677-1685. doi:10.1016/j.jacr.2024.02.034

87. The 2025 AI Index Report | Stanford HAI. Accessed October 11, 2025. https://hai.stanford.edu/ai-index/2025-ai-index-report

88. Parikh RB, Helmchen LA. Paying for artificial intelligence in medicine. *Npj Digit Med*. 2022;5(1):63. doi:10.1038/s41746-022-00609-6

89. amorse. Adoption of Artificial Intelligence in Healthcare Delivery Systems: Early Applications and Impacts. Peterson Health Technology Institute. March 25, 2025. Accessed October 11, 2025. https://phti.org/ai-adoption-early-applications-impacts/

90. Tardo L, Hirczy S, Blackburn K, et al. Evaluation of the impact of neurology electronic consults (e-consults): Experiences of a neurology resident clinic in a safety-net hospital. *J Telemed Telecare*. 2025;31(9):1270-1277. doi:10.1177/1357633X241273051

91. Bedi S, Liu Y, Orr-Ewing L, et al. Testing and Evaluation of Health Care Applications of Large Language Models: A Systematic Review. *JAMA*. Published online October 15, 2024. doi:10.1001/jama.2024.21700

92. *Health Data, Technology, and Interoperability: Certification Program Updates, Algorithm Transparency, and Information Sharing*. Vol 89. Federal Information & News Dispatch, LLC;

2024:1192. Accessed September 20, 2024. https://www.proquest.com/docview/2912085944/citation/14F0AB28E9DD4318PQ/1

93. Goh E, Gallo R, Hom J, et al. Large Language Model Influence on Diagnostic Reasoning: A Randomized Clinical Trial. *JAMA Netw Open*. 2024;7(10):e2440969. doi:10.1001/jamanetworkopen.2024.40969

94. Jabbour S, Fouhey D, Shepard S, et al. Measuring the Impact of AI in the Diagnosis of Hospitalized Patients: A Randomized Clinical Vignette Survey Study. *JAMA*. 2023;330(23):2275-2284. doi:10.1001/jama.2023.22295

95. Fine-Tuned ‘Small’ LLMs (Still) Significantly Outperform Zero-Shot Generative AI Models in Text Classification. Accessed December 11, 2024. https://arxiv.org/html/2406.08660v1

96. Tonmoy SMTI, Zaman SMM, Jain V, et al. A Comprehensive Survey of Hallucination Mitigation Techniques in Large Language Models. *arXiv*. Preprint posted online January 8, 2024. Accessed January 28, 2024. http://arxiv.org/abs/2401.01313

97. Open Medical-LLM Leaderboard - a Hugging Face Space by openlifescienceai. Accessed November 13, 2024. https://huggingface.co/spaces/openlifescienceai/open_medical_llm_leaderboard

98. Dettmers T, Pagnoni A, Holtzman A, Zettlemoyer L. QLoRA: Efficient Finetuning of Quantized LLMs. *arXiv*. Preprint posted online May 23, 2023. doi:10.48550/arXiv.2305.14314

99. Abadir PM, Battle A, Walston JD, Chellappa R. Enhancing Care for Older Adults and Dementia Patients With Large Language Models: Proceedings of the National Institute on Aging—Artificial Intelligence & Technology Collaboratory for Aging Research Symposium. *J Gerontol A Biol Sci Med Sci*. 2024;79(9):glae176. doi:10.1093/gerona/glae176

100. Obermeyer Z, Powers B, Vogeli C, Mullainathan S. Dissecting racial bias in an algorithm used to manage the health of populations. *Science*. 2019;366(6464):447-453. doi:10.1126/science.aax2342

101. Leslie D, Mazumder A, Peppin A, Wolters MK, Hagerty A. Does “AI” stand for augmenting inequality in the era of covid-19 healthcare? *The BMJ*. 2021;372:n304. doi:10.1136/bmj.n304

102. Savulescu J, Giubilini A, Vandersluis R, Mishra A. Ethics of artificial intelligence in medicine. *Singapore Med J*. 2024;65(3):150-158. doi:10.4103/singaporemedj.SMJ-2023-279

103. Protections (OHRP) O for HR. IRB Considerations on the Use of Artificial Intelligence in Human Subjects Research. October 21, 2022. Accessed September 23, 2024. https://www.hhs.gov/ohrp/sachrp-committee/recommendations/irb-considerations-use-artificial-intelligence-human-subjects-research/index.html

104. Sarkar AR, Chuang YS, Mohammed N, Jiang X. De-identification is not always enough. *arXiv*. Preprint posted online January 31, 2024. Accessed September 23, 2024. http://arxiv.org/abs/2402.00179

105. Broniatowski DA. *Psychological Foundations of Explainability and Interpretability in Artificial Intelligence*. National Institute of Standards and Technology (U.S.); 2021:NIST IR 8367. doi:10.6028/NIST.IR.8367

106. Tabassi E. *Artificial Intelligence Risk Management Framework (AI RMF 1.0)*. National Institute of Standards and Technology (U.S.); 2023:NIST AI 100-1. doi:10.6028/NIST.AI.100-1

107. Phillips PJ, Hahn CA, Fontana PC, et al. *Four Principles of Explainable Artificial Intelligence*. National Institute of Standards and Technology (U.S.); 2021:NIST IR 8312. doi:10.6028/NIST.IR.8312

108. Sim I, Cassel C. The Ethics of Relational AI — Expanding and Implementing the Belmont Principles. *N Engl J Med*. 2024;391(3):193-196. doi:10.1056/NEJMp2314771

109. Mello MM, Cohen IG. Regulation of Health and Health Care Artificial Intelligence. *JAMA*. 2025;333(20):1769-1770. doi:10.1001/jama.2025.3308